\documentclass{PoS}

\title{Silicon Photomultiplier Camera for Schwarzschild-Couder Cherenkov Telescopes}

\ShortTitle{Silicon Photomultiplier Camera for Schwarzschild-Couder Cherenkov Telescopes}

\author{J. Vandenbroucke for the CTA Consortium\\
        Physics Department and Wisconsin IceCube Particle Astrophysics Center, University of Wisconsin, Madison, WI 53706, USA\\
        E-mail: \email{vandenbrouck@wisc.edu}}

\abstract{
The Cherenkov Telescope Array (CTA) is an atmospheric Cherenkov observatory that will image the cosmos in very-high-energy gamma rays.  CTA will study the highest-energy particle accelerators in the Universe and potentially confirm the particle nature of dark matter.  We have designed an innovative Schwarzschild-Couder telescope which uses two mirrors to achieve excellent optical performance across a wide field of view.  The small plate scale of the dual-mirror optics enables a compact camera which uses modern technology including silicon photomultipliers and the TARGET application-specific integrated circuit to read out a finely pixelated focal plane of 11,328 channels with modest weight, volume, cost, and power consumption.  The camera design is hierarchical and modular at each level, enabling robust construction, operation, and maintenance.  A prototype telescope is under construction and will be commissioned at the VERITAS site in Arizona.  An array of such telescopes will provide excellent angular resolution and sensitivity in the core energy range of CTA, from 100 GeV to 10 TeV.
}

\FullConference{Technology and Instrumentation in Particle Physics 2014, Amsterdam, the Netherlands}

\begin{document}

\section{Particle Astrophysics with the Cherenkov Telescope Array}


In the past decade, ground-based TeV gamma-ray astronomy with imaging atmospheric Cherenkov telescopes (IACTs) has advanced from a novel technique to a basic tool of particle astrophysics.  As of 2014, 149 sources have been detected.\footnote{http://tevcat.uchicago.edu}  IACTs detect the Cherenkov light produced by gamma-ray-induced air showers that develop in the atmosphere.  Each telescope has a matrix of focal-plane photodetectors and associated electronics that records an image of the air shower, serving as a camera.  The photodetectors must be sensitive enough to detect single photons and fast enough to discriminate the few-nanosecond flash from the steady night sky.  The current generation of instruments (H.E.S.S., MAGIC, and VERITAS) has demonstrated that stereoscopic imaging with multiple telescopes provides excellent signal reconstruction and background rejection.  The next step is to build a larger array, the Cherenkov Telescope Array (CTA), featuring dozens of telescopes~\cite{concept}.  CTA will detect point-like gamma-ray sources with an order of magnitude greater sensitivity than the current generation of instruments.

\begin{figure}[h]
\centering
\includegraphics*[width=10cm]{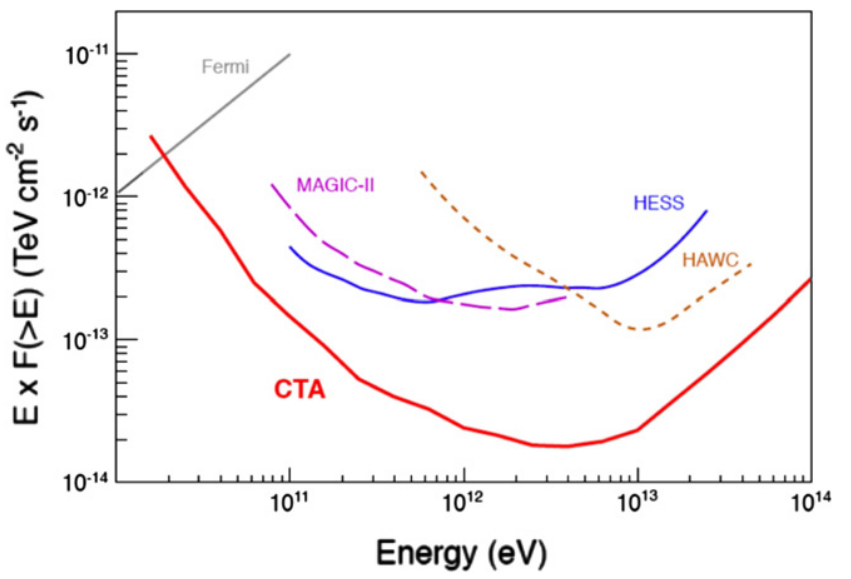}
\caption{Integral sensitivity of CTA to gamma-ray point sources, compared with other instruments.  Each IACT instrument is shown for a 50-hour observation.  Fermi and HAWC sensitivities are shown for one year of observations.  Figure from~\cite{concept}.}
\label{sensitivity}
\end{figure}

CTA will consist of both a Northern and a Southern array to provide full sky coverage.  The Southern array will feature four large (23~m diameter) telescopes with the light collecting power necessary to achieve a low (30~GeV) energy threshold; up to 25 single-mirror mid-size (12~m diameter) telescopes and up to 24 dual-mirror mid-size (9.5~m diameter) telescopes to cover the core energy range from 100 GeV to 10 TeV; and $\sim$70 small (4~m diameter) telescopes to span the 10~km$^2$ collecting area necessary for the lower gamma-ray fluxes detectable at several TeV and above.  The sensitivity of CTA for gamma-ray point sources is compared with other instruments in Figure~\ref{sensitivity}.

CTA will discover hundreds of new sources, each of which is an astrophysical particle accelerator comparable to or more powerful than the Large Hadron Collider.  Moreover, it will study the known TeV sources with much better sensitivity, sufficient to not only detect the sources but to study their morphology, spectral characteristics and temporal behavior.  Such quantitative studies are necessary to understand the mechanisms of astrophysical accelerators including supernova remnants, pulsars, pulsar wind nebulae, binary systems, active galactic nuclei, gamma-ray bursts, and likely new source classes that have not yet been detected at these extreme energies.  In addition to particle acceleration and cosmic-ray production, CTA will study dark matter.  By searching for gamma rays produced from dark matter annihilating or decaying \emph{in situ} in the cosmos (in gravitational wells such as the Galactic center and dwarf satellite galaxies of our Milky Way), CTA could confirm the particle nature of dark matter.  The observatory will complement the strong constraints produced by the Fermi Large Area Telescope, enabling a discovery or strong constraints for weakly interacting massive particles of higher mass than accessible to Fermi.  In addition to its gamma-ray measurements, CTA will be an excellent cosmic-ray electron and positron detector.  CTA will measure the total electron plus positron spectrum up to $\sim$100~TeV, and the charge-separated fluxes at the TeV scale using the Moon spectrometer method.  These measurements will provide important observational constraints on the many models developed to explain the surprising positron excess discovered over the past several years by PAMELA, Fermi, and AMS.

\section{Schwarzschild-Couder Telescopes}

Until now, IACTs have been built with single-mirror Davies-Cotton (DC) optics.  A superior optical design featuring dual-mirror Schwarzschild-Couder (SC) optics was proposed several years ago~\cite{vassiliev2007}.  While both designs typically use segmented mirrors, the dual-mirror technology features a secondary mirror in addition to the primary mirror.  The dual-mirror design enables both an excellent optical point-spread function across a wide field of view and a small camera plate scale.  The small camera plate scale results in a compact camera with modest weight, power consumption, and cost.  Moreover, the small plate scale enables modern photodetectors such as silicon photomultipliers (SiPMs) and readout electronics featuring application-specific integrated circuits that together provide high pixel density.  These features together result in telescope optics capable of finely focusing the Cherenkov light onto many small pixels, providing a high-resolution image of each air shower.  The improved resolution of the images improves the precision of gamma-ray direction and energy reconstruction and improves the ability to reject proton-induced background events, which are much more numerous than gamma-ray events.  The simulated performance of SC telescopes compared to DC telescopes is described in~\cite{jogler2013}.

\begin{figure}[h]
\centering
\includegraphics*[width=8cm]{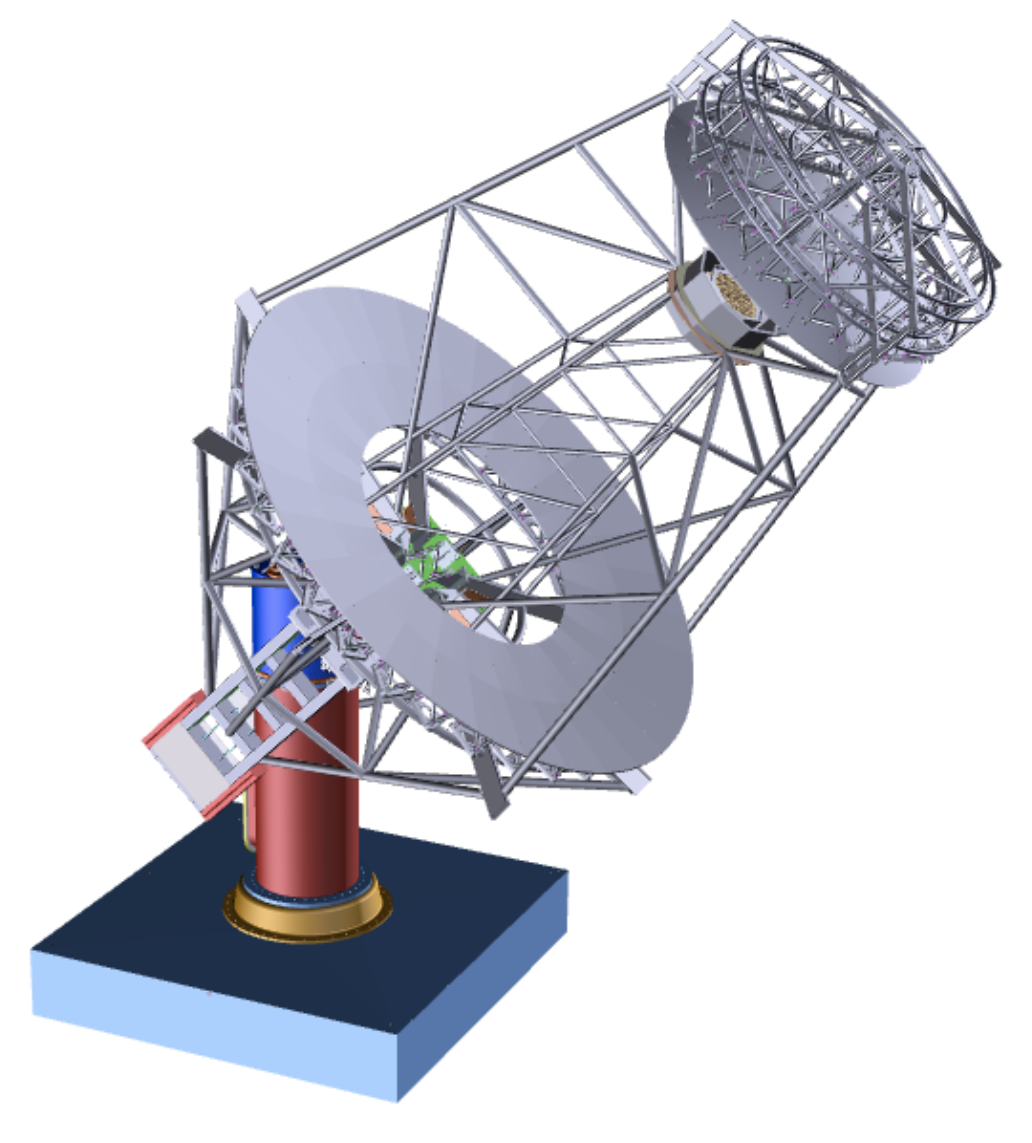}
\caption{Schwarzschild-Couder Telescope design.  The mount, primary mirror, secondary mirror, and camera are visible.}
\label{telescope}
\end{figure}

While much experience has been gained designing and operating Davies-Cotton telescopes over the past decade, a Schwarzschild-Couder telescope has not yet been built.  Prototype dual-mirror telescopes are now under construction for both the small-size and medium-size telescopes.  We focus here on the camera design for the medium-size Schwarzschild-Couder telescope (SCT).  Sub-system production is now underway for the prototype, which will be commissioned in 2015 at the VERITAS site in Arizona.  The prototype SCT will include a full primary mirror, a partial secondary mirror, full camera mechanical and positioning systems, and a partially populated focal plane and readout system.  The telescope design is shown in Figure~\ref{telescope}.  The optical design features a 9.5~m diameter primary mirror and an 8$^\circ$ diameter field of view.  In addition to the camera design for the medium-size SCT, a similar design (named the Compact High Energy Camera) is being developed for dual-mirror small-size telescopes.  The SCT and CHEC cameras share common components, including the front-end electronics and backplanes, as much as possible.

\section{Silicon Photomultiplier Camera Design}

While photomultiplier tubes (PMTs) have provided fast, sensitive photo-detection to particle physicists for decades, silicon photomultipliers (also known as Geiger-mode Avalanche Photodiodes or GAPDs) have become a preferred sensor for a growing number of applications.  They are mechanically robust, operate with relatively low voltage ($\sim$70~V), are not sensitive to magnetic fields, and can achieve a multiplication factor comparable to PMTs (10$^6$).  Their main disadvantages are that they have relatively high noise (dark current), relatively small collecting area, and temperature-dependent gain.  However, it was recognized several years ago that these are not serious disadvantages for IACTs because they operate in the night sky which has significant background light that is typically higher than the dark current, they use focusing optics so that a relatively small collecting area is sufficient (particularly with dual-mirror optics), and the temperature can either be stabilized or monitored to calibrate the gain, potentially with real-time bias voltage feedback to stabilize the gain.  Silicon photomultiplier technology is advancing rapidly, with several manufacturers introducing design improvements every few months.  A small single-mirror telescope featuring the first camera built with SiPMs for an IACT, FACT (First GAPD Cherenkov Telescope), has operated since 2011 on La Palma in the Canary Islands~\cite{FACT}.

\begin{figure}[h]
\centering
\includegraphics*[width=15cm]{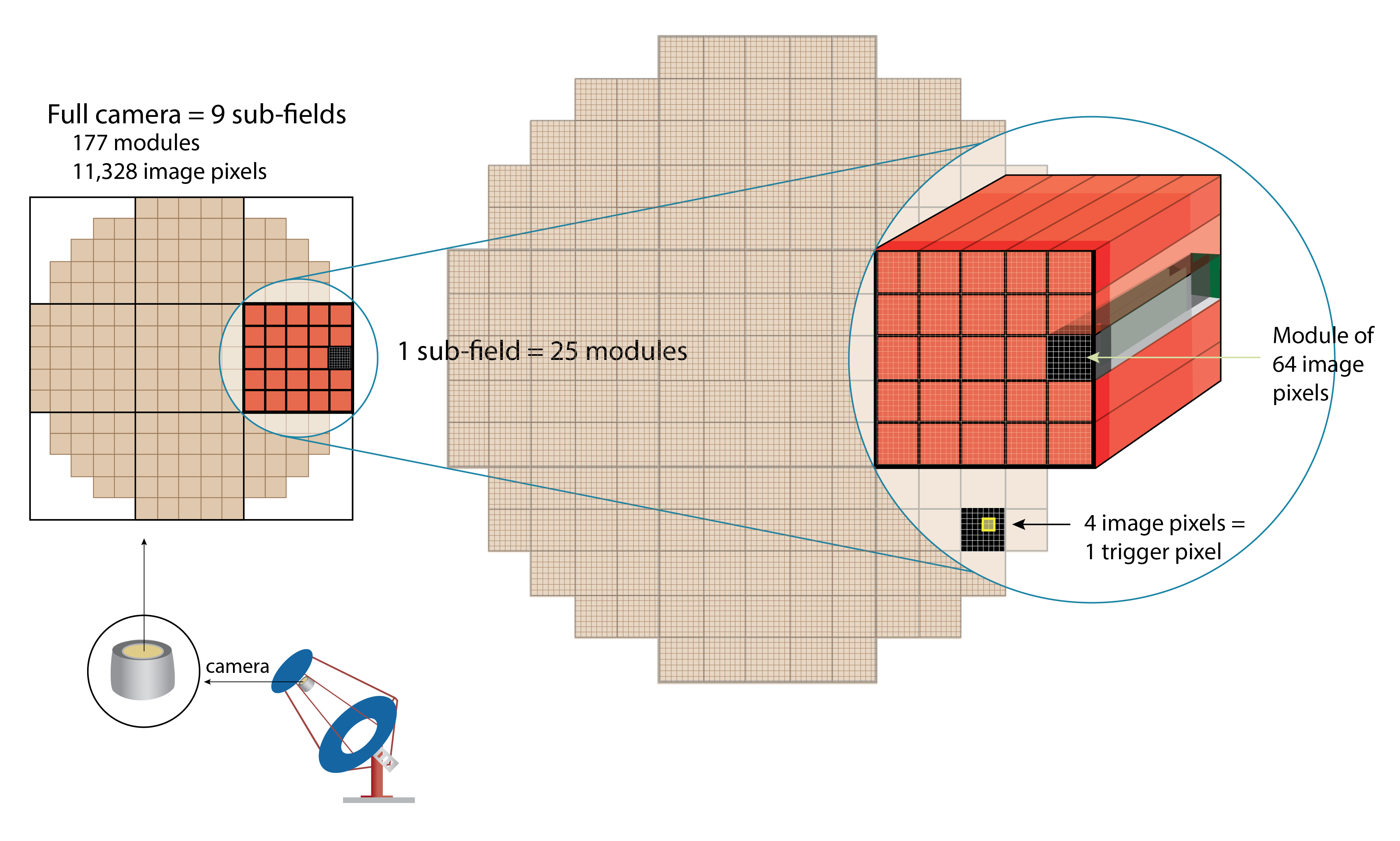}
\caption{Hierarchical, modular camera design.  A complete camera for a single telescope is composed of nine sub-fields, 177 camera modules, 2,832 trigger pixels, and 11,328 image pixels.}
\label{focal_plane}
\end{figure}


Because of the advantages described above, the SCT camera design is based on silicon photomultipliers.  In order to produce, operate, and maintain many telescopes each with many channels, we follow a design strategy that is hierarchical and modular at each level, as shown in Figure~\ref{focal_plane}.  The full focal plane consists of nine \emph{sub-fields}.  In order to approximate the circular focal plane, five of these are fully populated (each with 25 \emph{modules} arrayed in a square), while the four corner sub-fields are partially populated (each with 13 modules), for a total of 177 modules.  Each module consists of 64 pixels arrayed in a square, for a total of 11,328 pixels per telescope.  Each module consists of both a focal plane module and a front-end electronics module.  Each sub-field has a single backplane which connects to the front-end modules of the sub-field.  

The prototype SCT will include all components (SiPMs, focal plane modules, front-end electronics modules, and a backplane) for a complete sub-field.  It will be possible to operate the telescope with this sub-field in any of the nine positions to test the telescope optics.  We have characterized several SiPM models for CTA~\cite{bouvier2013}.  For the first sub-field of the prototype telescope, the SiPM model we selected is Hamamatsu S12642-0404PA-50(X).  This model features through-silicon-via technology to achieve low dead space.  Each device (\emph{tile}) contains 16 individual SiPM chips arrayed in a square.  Each chip is 3~mm $\times$ 3~mm.  We wire the outputs of each square of four chips together in parallel to form four \emph{image pixels} per tile, each 6~mm $\times$ 6~mm.  Each image pixel views 0.067$^\circ$ $\times$ 0.067$^\circ$ on the sky.

\begin{figure}[h]
\centering
\includegraphics*[height=5cm,trim = 6cm 0 8cm 0, clip]{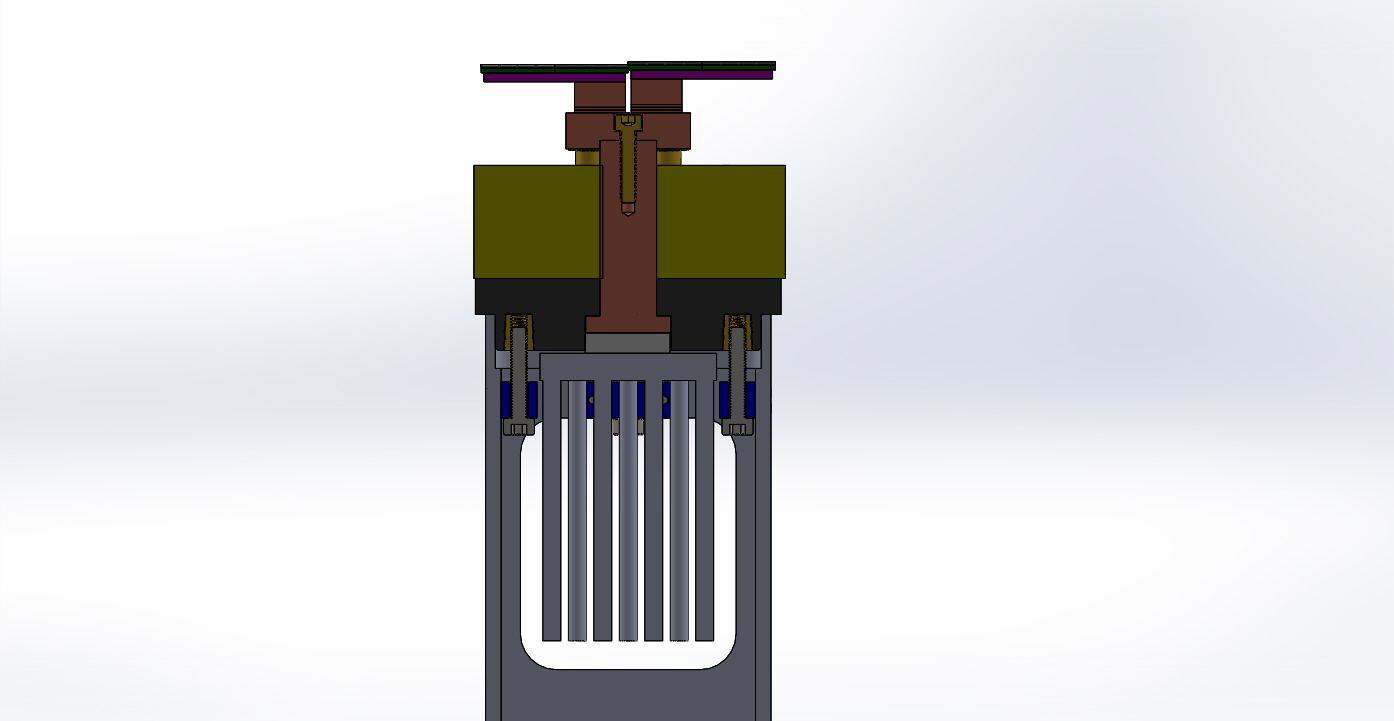}
\includegraphics*[height=5cm,trim = 5cm 0 9cm 0, clip]{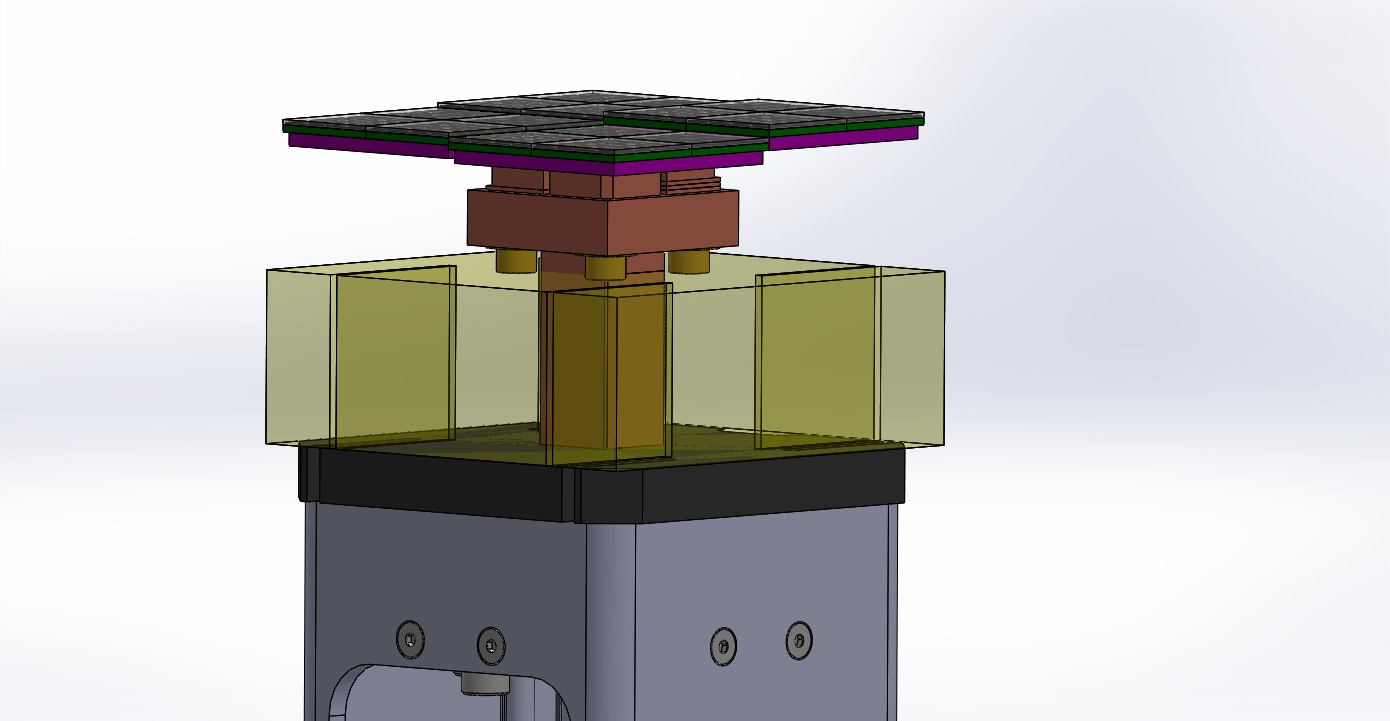}
\caption{Focal plane module from two perspectives.  The module contains a heat sink and copper finger with a Peltier element between them.  The module contains four quadrants, each with four SiPM tiles mounted on carrier boards.  Each SiPM tile of 16 chips provides one trigger pixel and four image pixels.  Insulating foam (yellow) surrounds the copper finger.}
\label{fpm}
\end{figure}

The silicon photomultipliers will be temperature stabilized at a temperature (between 0 and 20~$^\circ$C) below ambient.  The temperature stabilization, mechanical registration, and electrical connections for the SiPMs are provided by the focal plane modules (Figure~\ref{fpm}).  Temperature stabilization of each SiPM within $\pm$0.25~$^\circ$C across the focal plane and over time is achieved with a Peltier element and a copper heat conductor coupling the SiPM to the Peltier element.  Laboratory tests of the Peltier system have achieved better than $\pm$0.25~$^\circ$C stabilization.  All focal plane and front-end electronics modules are aligned parallel to the optical ($z$) axis.  To closely approximate the curvature of the ideal focal surface, the focal plane modules are offset relative to one another in $z$ and furthermore each focal plane module is divided into four \emph{quadrants} each of which is offset relative to one another in $z$.  This is achieved by machining the copper finger of each and then adjusting finely with shims.  Each focal plane module has 64 image pixels.  Each quadrant has 4 tiles.

The bias voltage supplied to each image pixel is the sum of a fixed 70 V voltage (provided by power supplies located behind the camera and routed through the backplanes) plus a trim voltage (provided by an analog-to-digital converter on the front-end electronics module) that can be adjusted between 0 and 4 V independently for each image pixel.  The operating voltage of each image pixel will be determined in order to operate all channels at a fixed gain.  In bright conditions such as operating in partial moonlight, the gain can be decreased.

Each focal plane module is connected to a front-end electronics module.  Each front-end electronics module features four TARGET (TeV Array Readout with GSa/s sampling and Event Trigger) chips~\cite{bechtol2012}.  TARGET chips are self-triggering switched capacitor array digitizers.  Each chip samples 16 channels in parallel, each with a deep (16,384 samples) analog buffer.  Cells from the analog buffer are digitized when a readout command is issued based on the TARGET trigger.  Triggering within TARGET is performed on the analog sum of four channels such that each \emph{trigger pixel} is composed of 2 $\times$ 2 image pixels.  The chip generates a low-voltage differential signaling (LVDS) trigger output with a width of several nanoseconds that is used by higher level trigger logic to make a readout decision.  In addition to the four TARGET chips, each front-end module has a single field-programmable gate array (FPGA) which controls the TARGET chips and receives their data, a control system for the Peltier elements, and a pre-amplifier circuit for each input analog channel between the SiPMs and the TARGET chips.  The pre-amplifiers feature two stages of gain for up to 100$\times$ total amplification, as well as a passive high-pass filter with a pole-zero cancellation resistor to shape the SiPM pulse.  While the SiPM output pulse has a fast rise and slow fall for an overall $\sim$40~ns full-width at half maximum (FWHM), the pre-amplifier output pulse has a FWHM of $\sim$10~ns.  By narrowing the pulse shape, we can determine the charge of each image pixel over a narrower integration window, reducing the contribution from electronic noise and night-sky background.

All 64 modules per sub-field connect to a single backplane board.  The backplane has an FPGA which performs the second level of trigger logic (\emph{camera trigger}).  If three contiguous trigger pixels pass the first level (TARGET) trigger in a short time window, the second level trigger is satisfied.  Both adjacent and diagonal trigger pixels count as contiguous.  Because there is no cross-communication between backplanes, the trigger pattern must be satisfied within a single sub-field.  Because the camera trigger logic is implemented in FPGAs, the coincidence time, multiplicity, and overall algorithm are programmable.  In addition to providing the camera trigger, the backplanes relay data, monitoring streams, and slow control commands between the front-end modules and the telescope data acquisition computer.

A Distributed Intelligent Array Trigger (DIAT) performs coincidence between neighboring telescopes in the array.  Because DIAT hardware is located at each telescope, this array trigger is decentralized rather than requiring all telescopes to report to a central location.  This array trigger logic is again implemented in FPGA and will be programmable.  In addition to pixel hit locations and times, the DIAT trigger will receive basic topological information such as image moments from each camera, which can be used as part of the array trigger decision.  The deep (16~$\mu$s) TARGET buffer provides sufficient time for cable propagation and array trigger logic to return a readout decision to the front-end modules.

\begin{figure}[h]
\centering
\includegraphics*[width=6cm]{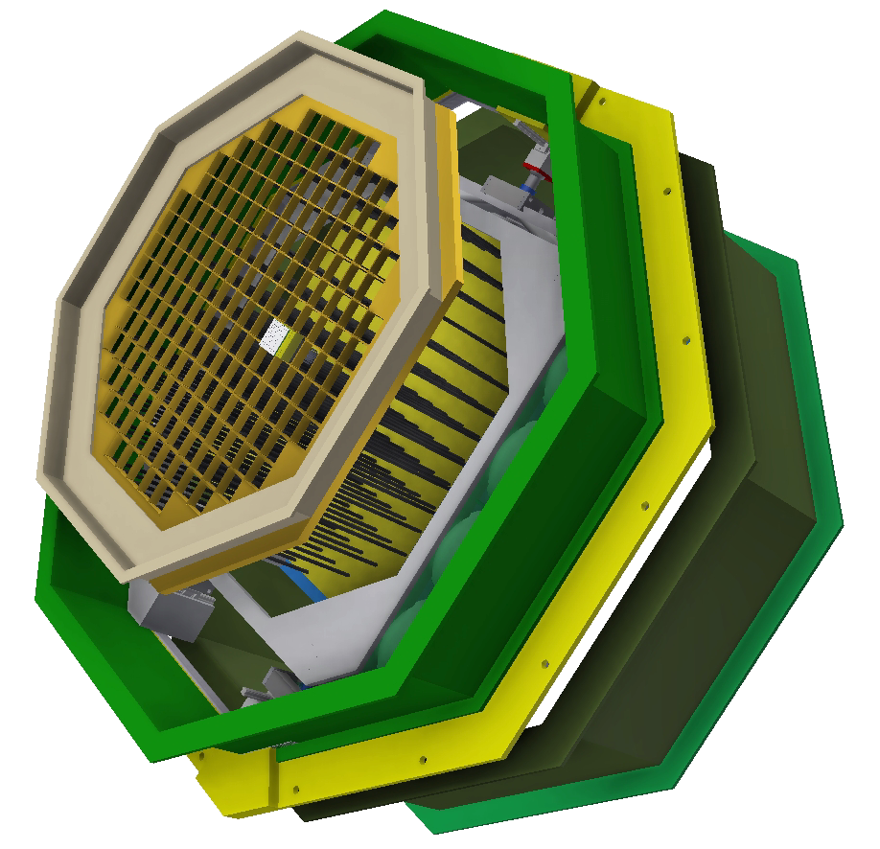}
\caption{Mechanical structure of the camera.  Camera modules are inserted in the lattice.}
\label{camera_full}
\end{figure}

The mechanical structure of the camera is shown in Figure~\ref{camera_full}.  A rack and cage system provides rigid support for the camera modules to meet the stringent alignment requirements of the telescope optics.  The rack is built with carbon fiber rods featuring low mass and low thermal expansivity.  Each camera module is mounted within an aluminum cage that slides into the rack from the front and attaches to an aluminum bulkhead at the back.  Connectors pass through slots in the bulkhead to mate with backplanes on the other side.  The mechanical system enables low focal plane deformation (less than 50~$\mu$m) under gravitational loads experienced for the full range of zenith angles that will be observed.

An active cooling system in the camera removes waste heat from the electronics and prevents thermal gradients that can cause focal plane deformations.  The cooling system uses high-flow fans blowing across a two-layer heat exchanger.  Thermal modeling indicates that deformations within 100~$\mu$m will be achieved.


An entrance window (6 mm thick polymethyl methacrylate) covering the SiPMs will protect them from the elements.  The window has good optical transmittance across the wavelength range detectable by the SiPMs, for the wide range of incidence angles (up to 60$^\circ$) that are reflected to the camera from the secondary mirror.  The peak power consumption of a full camera is estimated to be 7~kW.  The total mass of a full camera including  modules, backplanes, and mechanical structures is estimated to be 700~kg.


\section{Conclusion and Outlook}

We have designed an IACT camera well suited to the fine imaging capability of Schwarzschild-Couder optics.  The camera provides fine pixels to match the excellent optical point spread function of the dual-mirror optics.  The camera has more pixels than all 11 cameras of H.E.S.S., MAGIC, and VERITAS combined, in a volume smaller than any one of them.  We achieve this high channel density with innovative technology including silicon photomultipliers and TARGET self-triggering digitizer chips.  Sub-system production for a prototype SCT including telescope structure, mirrors, and camera is now underway.  The prototype will be commissioned in 2015 at the VERITAS site in Arizona.

\bigskip

{\bf Acknowledgements}
This research is supported by grants from the U.S. National Science Foundation and the U.S. Department of Energy Office of Science and by funding from the University of California, the Georgia Institute of Technology, the University of Minnesota, the Smithsonian Institution, SLAC National Accelerator Laboratory, Stanford University, the University of Wisconsin, the Ministry of Education, Culture, Sports, Science, and Technology in Japan, and Nagoya University.  We gratefully acknowledge support from the agencies and organizations listed at http://www.cta-observatory.org.



\bibliographystyle{elsart-num}
\bibliography{tipp_proceeding}



\end{document}